# Symmetries and conservation laws in the Lagrangian picture of quantum hydrodynamics

Peter Holland

Green Templeton College
University of Oxford
Oxford OX2 6HG
England

peter.holland@gtc.ox.ac.uk

## 1. Introduction

The similarity in form between the two real equations implied by the single-body spin 0 Schrödinger equation in the position representation (wave mechanics) and the equations of fluid mechanics with potential flow in its Eulerian formulation was first pointed out by Madelung in 1926 [1]. In this analogy, the probability density is proportional to the fluid density and the phase of the wavefunction is a velocity potential. A novel feature of the quantum fluid is the appearance of quantum stresses, which are usually represented through the quantum potential. To achieve mathematical equivalence of the models, the hydrodynamic variables have to satisfy conditions inherited from the wavefunction. These in turn provide physical insight into the original conditions. For example, the single-valuedness requirement on the wavefunction corresponds to the appearance of quantized vortices in the fluid. The hydrodynamic model has inspired several computational advances driven especially by theoretical chemists (for a comprehensive review up to the early 1980s see [2] and for more recent developments see [3]).

     Madelung's approach was based on the Eulerian picture and no reference was made to the Lagrangian picture of hydrodynamics, which conceives the motion of a fluid in terms of a continuum of trajectories and potentially opens the route to a new class of computational schemes. Contemporaneous with Madelung's work, de Broglie was developing a theory of trajectories in quantum mechanics but at this historic point an unfortunate digression occurred that, in the context of quantum hydrodynamics, took nearly 80 years to fully rectify. In the 1920s the notion of trajectory in quantum theory became irretrievably embroiled in issues of interpretation and its potential value as a tool in physics, regardless of its interpretational provenance, was not examined. The trajectory was so unacceptable that it was essentially banished from quantal discourse until Bohm reintroduced it in 1952 [4]. However, this restoration was still construed in a context circumvented by interpretational disputes. The earliest reference to the Lagrangian picture in quantum hydrodynamics appears to be due to Takabayasi [5] who, in the course of an early paper exploring the ramifications of Bohm's 1952 papers, mentioned in a footnote that obtaining the trajectories from the wavefunction corresponds to the transition from the Eulerian to the Lagrangian picture, and that in order to obtain the trajectories 'directly' one should start from the equations of motion in the Lagrangian

form (see eq. (15) below), subject to the subsidiary condition that the velocity field is irrotational. Takabayasi did not elaborate on what he meant by obtaining the trajectories 'directly' and, to our knowledge, neither he nor any other writer remarked upon the footnote subsequently. It took a further 20 years for the idea of employing the trajectories in methods to solve the wave equation to emerge. Subsequently, and especially since around 2000, many workers have contributed to the development of approximation techniques that actively employ the trajectories alongside the Eulerian equations (for references see [3,6]). It appears that the first clear demonstration that a congruence of trajectories – computed independently of the wavefunction (only the initial wavefunction is needed) – may exhibit sufficient structure to provide an exact method to deduce the time-dependence of the wavefunction, thereby laying the foundation for an alternative picture of quantum mechanics, was given in 2005 [6]. The exact constructive method has since been extended to flows in an arbitrary-dimensional Riemannian manifold [7,8], which has proved fruitful as it includes, as special cases, the many-body system, the inclusion of an external vector potential, relativistic spin theory including spin ½ [9] and spin 1 electromagnetism [7], and quantum fields [8]. The method has also been extended to a multiphase-flow model of quantum evolution [10] and to second-order (in time) field theories [11].

The Lagrangian approach introduces a new conception of state into quantum mechanics, complementary to the wavefunction, i.e., the time-dependent *position of a particle*, $q_i(a,t)$, and a new degree of freedom, i.e., the *particle label*, $a_i$. More precisely, the state is represented by the collective motion of the continuum of fluid particles obtained by continuously varying the label. Thus, unlike the Madelung (Eulerian) formulation, the Lagrangian picture adds variables to the quantum formalism but none is singled out as special. As has been pointed out [6], this implies that the Lagrangian picture exhibits a new quantum symmetry, or gauge freedom, *viz*. a continuous particle-relabelling covariance group of the dynamics with respect to which the Eulerian variables (position, density and velocity) are invariant. This is an analogue of the classical symmetry that is connected with vorticity conservation [12-16]. The origin of the relabelling symmetry is that the deformation coefficients (derivatives of the current position with respect to the label) appear in the field equations only through the Jacobian. This is a characteristic feature of fluid mechanics that is not displayed in other continuum theories (such as elasticity [15]) and reinforces the hydrodynamic analogy.

In this paper we examine the relabelling symmetry as a component of a general investigation of symmetries and conservation laws in the Lagrangian picture of quantum hydrodynamics, emphasizing their relation with symmetries and conservation laws in the Eulerian picture (which, being just wave mechanics, are well known). The emphasis is on general principles; whether the results will aid computational work remains to be investigated. The fact that the quantum Lagrangian picture has a variational basis gives us an opportunity to explore these connections through Noether's (first) theorem, which we formulate in the Lagrangian language. Analogous work has been done in classical Galilean-covariant hydrodynamics but no sufficiently general treatment is available from which to draw ready-made formulas suitable for application to the quantum regime. This is partly because the latter displays in the quantum internal energy higher-order derivatives of the fields (the position coordinates) than are customarily considered. Alongside the infinite relabelling group, the 12-parameter kinematical covariance group of the Schrödinger equation is derived within the Noetherian approach. Two methods to connect with the Eulerian treatment are described. A point we wish to stress is that the role of label transformations extends beyond the class corresponding to Eulerian invariance. As an example, it is shown that the linear superposition of waves, a fundamental symmetry of the quantum Eulerian picture, can be generated by a deformation-dependent label transformation in the Lagrangian picture.

## 2. Deduction of Schrödinger's equation from the Lagrangian-picture trajectory theory

### 2.1 Lagrangian picture

We here summarize the constructive method of obtaining the wave equation from a deterministic particle model given in ref [6].

In the Lagrangian picture of a fluid the history of the system is encoded in the state variables $q_i(a,t)$ (the indices $i,j,k,… = 1,2,3$), the positions of all the distinct fluid elements at time $t$, each particle being distinguished by a continuously variable vector label $a_i$. The particle label may be chosen to be the initial position or more abstractly as a point in some continuum, e.g., it may be a colour in a continuous spectrum. In order to have the freedom to choose physical position as a label we shall always assume that the label is a vector in a three-dimensional (Euclidian) space. Since only one particle can occupy a point at each time this label exhausts its identification. For labelling purposes we do not need to give, for example, the initial velocity (this information is however necessary for a fully posed dynamics). The motion is *continuous* in that the mapping from $a$-space to $q$-space is single-valued and differentiable with respect to $a_i$ and $t$ to whatever order is necessary, and the inverse mapping $a_i(q,t)$ exists and has the same properties. These assumptions are in accord with the properties of the single-valued velocity field implied by quantum mechanics (i.e., the ratio of current to density). The entire set of motions for all $a_i$ is termed a *flow*. The vectors $q_i$ and $a_i$ are referred to the same set of Cartesian space axes but they may also be regarded as related to one another by a time-dependent coordinate transformation. To each particle there is associated an elementary volume whose mass is conserved by the flow. The whole is structured by an internal potential derived from the density that represents a certain kind of particle interaction, and each particle responds to the potential via a force whose action is described by a form of Newton's second law. For all these reasons, we may regard the model as providing a 'particle' picture.

Let $\rho_0(a)$ be the initial quantal probability density. In the hydrodynamic model, $\rho_0(a)$ is identified with the initial number density (which is normalized: $\int \rho_0(a)d^3a = 1$). Then, introducing a mass parameter $m$, the mass of an elementary volume $d^3a$ attached to the point $a_i$ is given by $m\rho_0(a)d^3a$. The significance of the parameter $m$, conventionally described as the 'mass of the quantum system', is that it is the total mass of the fluid, since $\int m\rho_0(a)d^3a = m$. In this picture, the conservation of the mass of a fluid element in the course of its motion is expressed through the relation

$$m\rho(a,t)d^3q(a,t) = m\rho_0(a)d^3a \tag{1}$$

or

$$\rho(a,t) = J^{-1}(a,t)\rho_0(a) \tag{2}$$

where $J$ is the Jacobian of the transformation between the two sets of coordinates:

$$J = \det(\partial q_i/\partial a_j) = \frac{1}{3!}\varepsilon_{ijk}\varepsilon_{lmn}\frac{\partial q_i}{\partial a_l}\frac{\partial q_j}{\partial a_m}\frac{\partial q_k}{\partial a_n}, \quad 0 < J < \infty. \tag{3}$$

Here $\varepsilon_{ijk}$ is the completely antisymmetric tensor with $\varepsilon_{123} = 1$ and summation over repeated indices is always assumed.

Let $V$ be the potential of an external (classical) conservative body force and $U$ the internal potential energy of the fluid due to interparticle interactions. We assume that the Lagrangian has the same form as in the classical theory of ideal fluids except for the functional dependence of $U$: this depends on $\rho(q)$ and its first derivatives, and hence from (2) on the second-order derivatives of $q_i$ with respect to $a_i$, and is independent of other variables such as entropy. The Lagrangian is then

$$\begin{aligned}L &= \int \ell(q, \partial q/\partial t, \partial q/\partial a, \partial^2 q/\partial a^2, t) d^3 a \\ &= \int \left[\tfrac{1}{2} m \rho_0(a)\frac{\partial q_i}{\partial t}\frac{\partial q_i}{\partial t} - \rho_0(a)U(\rho) - \rho_0(a)V(q(a),t)\right] d^3 a.\end{aligned} \tag{4}$$

Here $\rho_0(a)$ and $V$ are prescribed functions and we substitute for $\rho$ from (2). We assume that $\rho_0$ and its derivatives vanish at infinity, which ensures that the surface terms in the variational principle vanish.

It is the action of the conservative force derived from $U$ on the trajectories that represents the quantum effects in this theory. As we shall see, these effects are characterized by the following choice for $U$:

$$U = \frac{\hbar^2}{8m}\frac{1}{\rho^2}\frac{\partial \rho}{\partial q_i}\frac{\partial \rho}{\partial q_i} = \frac{\hbar^2}{8m}\frac{1}{\rho_0^2}J_{ij}J_{ik}\frac{\partial}{\partial a_j}\left(\frac{\rho_0}{J}\right)\frac{\partial}{\partial a_k}\left(\frac{\rho_0}{J}\right) \tag{5}$$

where we have substituted from (2) and used

$$\frac{\partial}{\partial q_i} = J^{-1}J_{ij}\frac{\partial}{\partial a_j} \tag{6}$$

where

$$J_{il} = \frac{\partial J}{\partial(\partial q_i/\partial a_l)} = \frac{1}{2}\varepsilon_{ijk}\varepsilon_{lmn}\frac{\partial q_j}{\partial a_m}\frac{\partial q_k}{\partial a_n} \tag{7}$$

is the cofactor of $\partial q_i/\partial a_l$. The latter satisfies

$$\frac{\partial q_k}{\partial a_j}J_{ki} = J\delta_{ij}. \tag{8}$$

Clearly, $U$ has a local dependence on $\rho$ and its derivatives, and the coordinates $q_i$ enter only through the deformation gradients $\partial q_i/\partial a_j$ and their derivatives with respect to $a_i$.

The Euler-Lagrange equations for the coordinates,

$$\frac{\partial}{\partial t}\frac{\partial L}{\partial(\partial q_i(a)/\partial t)} - \frac{\delta L}{\delta q_i(a)} = 0, \tag{9}$$

where

$$\frac{\delta L}{\delta q_i} = \frac{\partial l}{\partial q_i} - \frac{\partial}{\partial a_j}\frac{\partial l}{\partial(\partial q_i/\partial a_j)} + \frac{\partial^2}{\partial a_j \partial a_k}\frac{\partial l}{\partial(\partial^2 q_i/\partial a_j \partial a_k)} \tag{10}$$

give the equation of motion of the *a*th fluid particle due to interparticle forces and the external force:

$$m\rho_0(a)\frac{\partial^2 q_i(a)}{\partial t^2} = -\rho_0(a)\frac{\partial V}{\partial q_i} - J_{kj}\frac{\partial \sigma_{ik}}{\partial a_j} \tag{11}$$

where

$$\sigma_{ij} = \frac{\hbar^2}{4m}\left(\frac{1}{\rho}\frac{\partial \rho}{\partial q_i}\frac{\partial \rho}{\partial q_j} - \frac{\partial^2 \rho}{\partial q_i \partial q_j}\right) \tag{12}$$

is a symmetric stress tensor which has been written in simplified form in terms of the dependent variables using (2) and (6). This (second order in *t*, fourth order in $a_i$) local nonlinear partial differential equation is the principal analytical result of the quantum Lagrangian method. For we shall see that, from its solutions $q_i(a,t)$, subject to specification of $\partial q_{i0}/\partial t$, we may derive solutions to Schrödinger's equation. *Motion in quantum mechanics may be regarded as the unravelling of a time-dependent coordinate transformation, $q_i(a,t)$.*

To obtain a flow that is representative of quantum mechanics we need to restrict the initial conditions of (11) to those that correspond to what we shall term "quasi-potential" flow. This means that the initial velocity field is of the form (we introduce the mass factor for later convenience)

$$\frac{\partial q_{i0}}{\partial t} = \frac{1}{m}\frac{\partial S_0(a)}{\partial a_i} \tag{13}$$

but the flow is not irrotational everywhere because the potential $S_0(a)$ (the initial quantal phase) obeys the quantization condition

$$\oint_{C_0} \frac{\partial q_{i0}}{\partial t} da_i = \oint_{C_0} \frac{1}{m}\frac{\partial S_0(a)}{\partial a_i} da_i = \frac{nh}{m}, \quad n \in \mathbb{Z}, \tag{14}$$

where $C_0$ is a closed curve. If it exists, vorticity occurs in nodal regions where the density vanishes and it is assumed that $C_0$ passes through a region of 'good' fluid, where $\rho_0 \neq 0$. To show that these assumptions imply motion characteristic of quantum mechanics we first demonstrate that they are preserved by the dynamical equation. To this end, we use a method based on Weber's transformation applied to the law of motion (11) in its 'Lagrangian' form:

$$m\frac{\partial^2 q_i}{\partial t^2}\frac{\partial q_i}{\partial a_k} = -\frac{\partial}{\partial a_k}(V + V_Q) \tag{15}$$

where

$$V_Q = \frac{\hbar^2}{4m\rho}\left(\frac{1}{2\rho}\frac{\partial \rho}{\partial q_i}\frac{\partial \rho}{\partial q_i} - \frac{\partial^2 \rho}{\partial q_i \partial q_i}\right) \tag{16}$$

is the de Broglie-Bohm quantum potential [17]. Integrating this equation between the time limits $(0,t)$ and substituting (13) give

$$\frac{\partial q_i}{\partial t}\frac{\partial q_i}{\partial a_k} = \frac{1}{m}\frac{\partial S}{\partial a_k}, \quad S(a,t) = S_0(a) + \chi(a,t), \tag{17}$$

where

$$\chi(a,t) = \int_0^t \left(\tfrac{1}{2}m\left(\frac{\partial q_i}{\partial t}\right)^2 - V - V_Q\right)dt \tag{18}$$

with initial conditions $q_{i0} = a_i$, $\chi_0 = 0$. The left-hand side of (17) gives the velocity at time $t$ with respect to the $a$-coordinates and this is obviously a gradient. To obtain the $q$-components we multiply by $J^{-1}J_{ik}$ and use (6) and (8) to get

$$\frac{\partial q_i}{\partial t} = \frac{1}{m}\frac{\partial S}{\partial q_i} \tag{19}$$

where $S = S(a(q,t),t)$. Thus, for all time, the velocity of each particle is the gradient of a potential with respect to the current position.

To complete the demonstration, we note that the motion is quasi-potential since the value (14) of the circulation is preserved following the flow:

$$\frac{\partial}{\partial t}\oint_C \frac{\partial q_i}{\partial t}dq_i = 0 \tag{20}$$

where $C$ is a curve moving with the flow. This theorem has been stated previously in the quantum context [18] and will be rederived below using relabelling invariance. We conclude that each particle retains forever the quasi-potential property if it possesses it at any moment.

**2.2 Eulerian picture**

The fundamental link between the particle (Lagrangian) and wave-mechanical (Eulerian) pictures is defined by the following expression for the Eulerian density:

$$\rho(x,t) = \int \delta(x - q(a,t)) \rho_0(a) d^3a. \qquad (21)$$

The corresponding formula for the Eulerian velocity is contained in the expression for the current:

$$\rho(x,t) v_i(x,t) = \int \frac{\partial q_i(a,t)}{\partial t} \delta(x - q(a,t)) \rho_0(a) d^3a. \qquad (22)$$

Evaluating the integrals, (21) and (22) are equivalent to the following local expressions

$$\rho(x,t) = J^{-1}\big|_{a(x,t)} \rho_0(a(x,t)), \qquad (23)$$

$$v_i(x,t) = \frac{\partial q_i(a,t)}{\partial t}\bigg|_{a(x,t)}. \qquad (24)$$

Relation (23) restates the conservation equation (2), and (24) gives the relation between the velocities in the two pictures.
   These formulas enable us to translate the Lagrangian flow equations into Eulerian language. Differentiating (21) with respect to $t$ and using (22) we easily deduce the continuity equation

$$\frac{\partial \rho}{\partial t} + \frac{\partial}{\partial x_i}(\rho v_i) = 0. \qquad (25)$$

Next, differentiating (22) and using (15) and (25), we get the quantum analogue of Euler's equation:

$$\frac{\partial v_i}{\partial t} + v_j \frac{\partial v_i}{\partial x_j} = -\frac{1}{m} \frac{\partial}{\partial x_i}(V + V_Q). \qquad (26)$$

Finally, the quasi-potential condition (19) becomes

$$v_i = \frac{1}{m} \frac{\partial S(x,t)}{\partial x_i}. \qquad (27)$$

The formulas (23) and (24) give the general solution of the continuity equation (25) and Euler equation (26) in terms of the trajectories and the initial density.
   To establish the connection between the Eulerian equations and Schrödinger's equation, it is a simple matter to deduce from (26) and (27) that

$$\frac{\partial S}{\partial t} + \frac{1}{2m} \frac{\partial S}{\partial x_i} \frac{\partial S}{\partial x_i} + V + V_Q = 0 \qquad (28)$$

where we have absorbed a function of $t$ in $S$. Combining (25), (27) and (28), the function $\psi(x,t) = \sqrt{\rho}\exp(iS/\hbar)$ obeys Schrödinger's equation:

$$i\hbar \frac{\partial \psi}{\partial t} = -\frac{\hbar^2}{2m}\frac{\partial^2 \psi}{\partial x_i \partial x_i} + V\psi. \qquad (29)$$

We have deduced this from the Lagrangian particle equation (11) subject to the quasi-potential requirement.

This procedure enables us to write down an explicit formula for the time-dependent wavefunction in terms of the trajectories, up to a global phase, given the initial wavefunction $\psi_0(a) = \sqrt{\rho_0}\exp(iS_0/\hbar)$. First, solve (11) subject to the initial conditions $q_{i0}(a) = a_i, \partial q_{i0}/\partial t = m^{-1}\partial S_0/\partial a_i$ to get the ensemble of trajectories for all $a_i, t$. Next, substitute $q_i(a,t)$ in (23) to find $\rho$ and $\partial q_i/\partial t$ in (24) to get $\partial S/\partial x_i$. This gives $S$ up to an additive function of time $f(t)$. To fix this function, apart from an additive constant, use (28). We obtain finally for the wavefunction

$$\psi(x,t) = \sqrt{\left(J^{-1}\rho_0\right)\big|_{a(x,t)}} \exp\left[\frac{i}{\hbar}\left(\int m\partial q_i/\partial t\big|_{a(x,t)} dx_i + f(t)\right)\right]. \qquad (30)$$

Note that this method of solution does not dispense with the wavefunction – the initial form of the latter is integral to the dynamical equation of the trajectories (through the density) and to its initial conditions (through the phase).

The Eulerian equations (25) and (26) form a closed system of four first-order coupled partial differential equations to determine the four independent 'basic' fields $\rho(x)$, $v_i(x)$. The erasure from them of the particle variables is part of the reason why the Eulerian language is particularly suited to represent the wave-mechanical formalism, which likewise, of course, makes no reference to the trajectory concept. The passage from the Lagrangian to the Eulerian picture is a reductive process in which the number of dependent variables decreases [19]. In this connection, we note that any relation between Lagrangian variables can be recast formally in the Eulerian *language* by writing $a_i(x,t)$ but it will not always be a statement in the Eulerian *picture* since the derived functions of $x_i$ and $t$ may not be reducible to the basic Eulerian set $\rho$, $v$ and their derivatives. We find this in the canonical phase space formulation of the theory where extraneous advected variables appear [6] and in the example of Sec. 5 (cf. eq. (60)).

**2.3 Equivalence of first-order and second-order trajectory laws**

We have established above that the second-order trajectory equation (11) implies the first-order law (17) where the potential $S$ is determined by (18). The converse is trivially proved: given (17) and (18) we can deduce (11) by differentiation. Hence, the first- and second-order formulations are formally *equivalent*. Each may have advantages where the other is deficient. For example, if we wish to compute the trajectories, knowing only $\psi_0$ and without invoking the Eulerian equations as aids, the second-order version must be used. On the other hand, if we desire to compute the trajectories from a known wavefunction the first-order version may be preferable. It would artificially restrict the insights and opportunities afforded by quantum hydrodynamics to treat one law as more fundamental than the other, either conceptually or computationally.

## 3. Eulerian identity transformation

We commence our analysis of symmetries in quantum hydrodynamics by establishing a simple but fundamental result. A continuous transformation of the independent and dependent variables in one picture, in particular a symmetry (a transformation that leaves the dynamical equations covariant), will have a continuous image in the other picture. However, this correspondence will not be 1-1. Since the Eulerian picture is a reduction of the Lagrangian one, a unique transformation of the Eulerian variables generally corresponds to a *class* of transformations of the Lagrangian variables. More specifically, *a Lagrangian-picture symmetry corresponding to an Eulerian-picture symmetry will be unique only up to a relabelling transformation, independent of the parameters defining the Eulerian transformation.*

To demonstrate this property it is sufficient to examine the identity transformation in the Eulerian picture, which is, for any fluid described by the density and velocity fields,

$$x_i' = x_i, \quad t' = t, \quad v_i'(x',t') = v_i(x,t), \quad \rho'(x',t') = \rho(x,t). \tag{31}$$

This corresponds to the following transformation of the Lagrangian variables:

$$q_i'(a',t') = q_i(a,t), \quad t' = t, \quad \frac{\partial q_i'}{\partial t'} = \frac{\partial q_i}{\partial t}, \quad \rho_0'(a')J'^{-1} = \rho_0(a)J^{-1} \tag{32}$$

where $a_i' = a_i'(a,t)$. The latter is therefore not generally an identity transformation. To discover the function $a_i'$, we write $d = \det\left(\partial(t',a_i')/\partial(t,a_j)\right) \neq 0$ and $D = \det\left(\partial a_i'/\partial a_j\right) \neq 0$, with $d(t'=t) = D$, and use the first two members of (32) to get

$$\frac{\partial q_i'}{\partial a_j'} = \frac{1}{D}\frac{\partial d}{\partial\left(\partial a_j'/\partial a_k\right)}\bigg|_{t'=t}\frac{\partial q_i}{\partial a_k}, \quad \frac{\partial q_i'}{\partial t'} = \frac{\partial q_i}{\partial t} + \frac{1}{D}\frac{\partial d}{\partial\left(\partial t'/\partial a_j\right)}\bigg|_{t'=t}\frac{\partial q_i}{\partial a_j}. \tag{33}$$

Comparing with the last two members of (32) we obtain

$$\frac{\partial a_i'}{\partial t} = 0, \quad \rho_0'(a')D = \rho_0(a). \tag{34}$$

Thus, the corresponding transformation in the Lagrangian picture constitutes a time-independent diffeomorphism $a_i'(a)$, or relabelling of the fluid particles, with respect to which the reference density transforms as a tensor density. Invariants of the transformation include the number of particles in an elementary volume of label space, i.e.,

$$\rho_0'(a')d^3a' = \rho_0(a)d^3a, \tag{35}$$

and their position and velocity. The relabelling is arbitrary if no other conditions are required on the transformation of $\rho_0$. Otherwise, (34) may involve a constraint on the relabelling (see Sec. 5).

A Lagrangian-picture theory that is reducible to the basic Eulerian variables will be covariant with respect to the transformation (34). The Lagrangian (4) can be written as $L = \int \left[ \tfrac{1}{2} m\rho v^2 - \rho U(\rho, \partial\rho/\partial x) - \rho V(x) \right] d^3x$ and hence falls into this category (note that this Lagrangian cannot be used in a variational principle to derive (25) and (26) as Euler-Lagrange equations). Label transformations other than the class (34) do not generally leave the Eulerian functions invariant.

As mentioned previously, a common choice for labelling is the initial particle position: $q_{i0}(a) = a_i$. This is intuitively reasonable and implies that the element $d^3a$ coincides with an elementary spatial volume. Since, according to (32), $q'_i(a', t' = 0) = q_i(a, t = 0)$ and, in general, $a_i \neq a'_i$, the transformation (34) expresses the freedom to choose a label other than the initial position. A natural choice is to distort the coordinates $a_i$ so that the density is uniform with respect to them: $\rho'_0(a') = k = $ constant [11]. In one dimension this is achieved by the labelling

$$a' = k^{-1} \int_{-\infty}^{a} \rho_0(a) da. \tag{36}$$

Starting from an arbitrary labelling, we can transform to a labelling which coincides with initial position by observing that, in general, $q_i(a, t = 0) = f_i(a)$, so we just set $a'_i = f_i(a)$, requiring only that the Jacobian of the transformation is positive.

Some benefits of the label symmetry include a means to simplify the Lagrangian-picture problem and, as shown below, an effective way of generating a class of conservation laws.

**4. Noether's theorem in label space**

**4.1 Conservation in the Lagrangian picture**

A typical textbook illustration of Noether's first theorem employs a known continuous symmetry of a system of differential equations (the Euler-Lagrange equations deduced from a variational principle) that leaves the action functional invariant to derive an associated conserved charge. Actually, the theorem allows one to do more than this since the conservation laws it provides determine a class of symmetries of the differential equations (corresponding to the chosen Lagrangian), which need not then be known in advance. The class of symmetry transformations so obtained will not generally be exhaustive of all the continuous symmetries admitted by the dynamical equations since some dynamical symmetries may not be Noetherian symmetries with respect to the chosen Lagrangian, and those obtained are contingent on that choice. For these reasons, the implied set of conserved charges may likewise not be comprehensive. Nevertheless, as we shall see in our example, a sufficiently broad class of symmetries and associated conserved charges including all the principal expected ones can be generated by a straightforward application of this method.

This approach – of deriving rather than assuming the transformation functions – has been pursued previously in a similar context, that of the Lagrangian picture of a classical ideal fluid [20]. We shall generalize the previous treatment to allow for the appearance of higher field derivatives in the internal potential energy, and for the external potential, and will include symmetries that were missed in the cited prior work. A version of Noether's theorem general enough for our purposes follows (for more details see [21]).

Consider a Lie group of transformations of the independent and dependent variables:

$$t' = t + \varepsilon \xi_0(q,a,t), \quad a'_i = a_i + \varepsilon \xi_i(q,a,t), \quad q'_i(a',t') = q_i(a,t) + \varepsilon \eta_i(q,a,t), \tag{37}$$

where $\varepsilon$ is a dimensionless infinitesimal parameter. Note that the functional dependence of the transformation functions $\xi_0$, $\xi_i$ and $\eta_i$ on $a_i$ and $t$ may be explicit or implicit via $q_i(a,t)$. To take account of this dependence, we use the following notation for the derivatives with respect to the independent variables:

$$\left. \begin{array}{l} \dfrac{D}{\partial t} = \dfrac{\partial}{\partial t} + \dfrac{\partial q_i}{\partial t}\dfrac{\partial}{\partial q_i} + \dfrac{\partial^2 q_i}{\partial t^2}\dfrac{\partial}{\partial(\partial q_i/\partial t)} \\[2mm] \dfrac{D}{\partial a_i} = \dfrac{\partial}{\partial a_i} + \dfrac{\partial q_j}{\partial a_i}\dfrac{\partial}{\partial q_j} + \dfrac{\partial^2 q_j}{\partial a_i \partial a_k}\dfrac{\partial}{\partial(\partial q_j/\partial a_k)} + \ldots \end{array} \right\} \tag{38}$$

The induced infinitesimal transformations of the derivatives of the dependent variables are:

$$\left. \begin{array}{l} \dfrac{\partial q'_i}{\partial t'} = \dfrac{\partial q_i}{\partial t} + \varepsilon \dfrac{D\eta_i}{\partial t} - \varepsilon \dfrac{D\xi_0}{\partial t}\dfrac{\partial q_i}{\partial t} - \varepsilon \dfrac{D\xi_k}{\partial t}\dfrac{\partial q_i}{\partial a_k} \\[3mm] \dfrac{\partial q'_i}{\partial a'_j} = \dfrac{\partial q_i}{\partial a_j} + \varepsilon \dfrac{D\eta_i}{\partial a_j} - \varepsilon \dfrac{D\xi_0}{\partial a_j}\dfrac{\partial q_i}{\partial t} - \varepsilon \dfrac{D\xi_l}{\partial a_j}\dfrac{\partial q_i}{\partial a_l} \\[3mm] \dfrac{\partial^2 q'_i}{\partial a'_k \partial a'_j} = \dfrac{\partial^2 q_i}{\partial a_k \partial a_j} + \varepsilon \dfrac{D^2\eta_i}{\partial a_k \partial a_j} - \varepsilon \dfrac{D^2\xi_0}{\partial a_k \partial a_j}\dfrac{\partial q_i}{\partial t} - \varepsilon \dfrac{D\xi_0}{\partial a_k}\dfrac{\partial^2 q_i}{\partial a_j \partial t} - \varepsilon \dfrac{D\xi_0}{\partial a_j}\dfrac{\partial^2 q_i}{\partial a_k \partial t} \\[3mm] \qquad\qquad - \varepsilon \dfrac{D^2\xi_l}{\partial a_k \partial a_j}\dfrac{\partial q_i}{\partial a_l} - \varepsilon \dfrac{D\xi_l}{\partial a_k}\dfrac{\partial^2 q_i}{\partial a_j \partial a_l} - \varepsilon \dfrac{D\xi_l}{\partial a_j}\dfrac{\partial^2 q_i}{\partial a_k \partial a_l}. \end{array} \right\} \tag{39}$$

The invariance of the action $\int \ell \, d^3a \, dt$ under the transformation (37) entails the local condition

$$\ell\bigl(q', \partial q'/\partial t', \partial q'/\partial a', \partial^2 q'/\partial a'^2, t', a'\bigr)\left(1 + \varepsilon \dfrac{D\xi_0}{\partial t} + \varepsilon \dfrac{D\xi_i}{\partial a_i}\right)$$
$$= \ell\bigl(q, \partial q/\partial t, \partial q/\partial a, \partial^2 q/\partial a^2, t, a\bigr) + \varepsilon \left(\dfrac{D\Lambda_0}{\partial t} + \dfrac{D\Lambda_i}{\partial a_i}\right) \tag{40}$$

where the functions $\Lambda_0, \Lambda_i$ depend on $t, a_i, q_i, \partial q_i/\partial a_j$. Expanding the left-hand side to order $\varepsilon$, rearranging and subjecting the fields $q_i$ to the Euler-Lagrange equations (9), Noether's theorem asserts that (40) takes the form of a continuity equation in *a-t* space:

$$\frac{D\mathrm{P}}{\partial t}+\frac{DJ_j}{\partial a_j}=0, \tag{41}$$

where the density and current are given by

$$\mathrm{P}=\ell\xi_0+\frac{\partial\ell}{\partial(\partial q_i/\partial t)}\left(\eta_i-\frac{\partial q_i}{\partial t}\xi_0-\frac{\partial q_i}{\partial a_l}\xi_l\right)-\Lambda_0 \tag{42}$$

$$J_j=\ell\xi_j+\left(\frac{\partial\ell}{\partial(\partial q_i/\partial a_j)}-\frac{D}{\partial a_k}\frac{\partial\ell}{\partial(\partial^2 q_i/\partial a_j \partial a_k)}\right)\left(\eta_i-\frac{\partial q_i}{\partial t}\xi_0-\frac{\partial q_i}{\partial a_l}\xi_l\right) \\ +\frac{\partial\ell}{\partial(\partial^2 q_i/\partial a_j \partial a_k)}\frac{D}{\partial a_k}\left(\eta_i-\frac{\partial q_i}{\partial t}\xi_0-\frac{\partial q_i}{\partial a_l}\xi_l\right)-\Lambda_j. \tag{43}$$

Then, invoking Gauss's theorem and assuming the fields vanish at infinity, we obtain a conservation law:

$$\frac{d}{dt}\int_{-\infty}^{\infty}\mathrm{P}(a,t)d^3a=0. \tag{44}$$

The continuity equation (41) both generates a set of conserved quantities and determines the class of transformation functions (37) and associated restrictions on $V$ that leave the action invariant, one charge being associated with each transformation function and constraint on $V$. To determine these functions we note that there is no functional relationship between the derivatives of $q_i$ beyond the Euler-Lagrange equations (11). Inserting the Lagrangian density $\ell$ from (4) in (41), we must therefore set to zero the coefficients of the independent derivatives of $q_i$ with respect to $a_i$ and $t$ and their products. The outcome of an arduous calculation is that the most general form of the transformation functions is the following:

$$\xi_0=d+\beta t+\alpha t^2, \quad \xi_i=\xi_i(a), \quad \eta_i=\left[(\tfrac{1}{2}\beta+\alpha t)\delta_{ij}+\omega_{ij}\right]q_j(a,t)-u_i t+c_i, \\ \Lambda_0=m\rho_0(a)(\tfrac{1}{2}\alpha q_i q_i-u_i q_i), \quad \Lambda_i=0 \tag{45}$$

where

$$\frac{\partial}{\partial a_i}(\rho_0\xi_i)=0, \tag{46}$$

$$\eta_i\frac{\partial V}{\partial q_i}+\xi_0\frac{\partial V}{\partial t}+V\frac{\partial\xi_0}{\partial t}=0, \tag{47}$$

and $\omega_{ij}=-\omega_{ji},u_i,c_i,d,\alpha,\beta$ are constants. Unimportant functions of $t$ and $a_i$ in $\Lambda_0$ and $\Lambda_i$ have been ignored. The constants, 12 in all, parameterize the group in addition to the function $\xi_i$, which is arbitrary save for the condition (46) in which $\rho_0$ is a prescribed function. A series of conservation laws follow by inserting specific values of the parameters in (42) and (43), as we shall see below. The absence of $\hbar$ in these transformation formulas suggests that they have a non-quantum origin. This turns out to be the case: they, in fact, define the maximal kinematical symmetry group of the classical Hamilton-Jacobi equation (obtained when $V_Q$ is negligible in (28)).

**4.2 Conservation in the Eulerian picture**

One of the benefits of the Lagrangian-coordinate approach to continuum mechanics is that it provides an additional means to discover Eulerian conserved charges, particularly through the application of Noether's theorem. This is especially useful in cases where the Eulerian-picture transformation corresponding to a Lagrangian-picture symmetry is trivial. We can connect the descriptions of conservation in the two pictures in two ways.

First, we can convert the Lagrangian continuity equation (41) into a corresponding Eulerian one relating a density $\bar{P}(x,t)$ and current $\bar{J}_i(x,t)$,

$$\frac{D\bar{P}}{\partial t} + \frac{D\bar{J}_i}{\partial x_i} = 0, \tag{48}$$

via the conversion formulas [22-24]

$$\bar{P}(x,t) = P(a,t)J^{-1}(a,t)\Big|_{a(x,t)}, \quad \bar{J}_i(x,t) = \left(PJ^{-1}\frac{\partial q_i}{\partial t} + \frac{\partial q_i}{\partial a_j}J_j\right)\Bigg|_{a(x,t)}. \tag{49}$$

We may thus deduce from the Lagrangian conservation law an Eulerian conservation law:

$$\frac{d}{dt}\int_{-\infty}^{\infty}\bar{P}(x,t)d^3x = 0. \tag{50}$$

As an example, we consider the Lagrangian density $P = \rho_0$, which obeys the equation $DP/\partial t = 0$ with $J_i = 0$. Then, from (49), $\bar{P} = \rho, \bar{J}_i = \rho v_i$, and (48) is just (25).

A second way to connect the Lagrangian and Eulerian accounts of conservation is to compare (50) with the conserved charge obtained directly in the Eulerian formulation using the symmetry transformation that corresponds to (45). With reference to the standard Lagrangian density for the Schrödinger field,

$$\hat{\ell} = \frac{i\hbar}{2}\left(\psi^*\frac{\partial\psi}{\partial t} - \psi\frac{\partial\psi^*}{\partial t}\right) - \frac{\hbar^2}{2m}\frac{\partial\psi^*}{\partial x_i}\frac{\partial\psi}{\partial x_i} - V\psi^*\psi, \tag{51}$$

we consider the infinitesimal transformation

$$t' = t + \varepsilon\theta_0(x,t), \quad x'_i = x_i + \varepsilon\theta_i(x,t), \quad \psi'(x',t') = \psi(x,t) + \varepsilon\phi(x,t),$$
$$\psi'^*(x',t') = \psi^*(x,t) + \varepsilon\phi^*(x,t). \tag{52}$$

The conserved density and current implied by Noether's theorem, which obey (48), are

$$\hat{P} = \hat{\ell}\theta_0 + \left[\frac{i\hbar}{2}\psi^*\left(\phi - \frac{\partial\psi}{\partial t}\theta_0 - \frac{\partial\psi}{\partial x_i}\theta_i\right) + \text{cc}\right] - \hat{\Lambda}_0 \tag{53}$$

$$\hat{J}_j = \hat{\ell}\theta_j - \left[\frac{\hbar^2}{2m}\frac{\partial\psi^*}{\partial x_j}\left(\phi - \frac{\partial\psi}{\partial t}\theta_0 - \frac{\partial\psi}{\partial x_i}\theta_i\right) + \text{cc}\right] - \hat{\Lambda}_j. \tag{54}$$

The transformation functions (52) corresponding to (45) are

$$\theta_0 = d + \beta t + \alpha t^2, \quad \theta_i = \left(\tfrac{1}{2}\beta + \alpha t\right)x_i + \omega_{ij}x_j - u_i t + c_i,$$

$$\phi = \psi\left[-\tfrac{3}{2}\left(\tfrac{1}{2}\beta + \alpha t\right) + \frac{im}{\hbar}\left(\tfrac{1}{2}\alpha x_i x_i - u_i x_i\right)\right], \quad \hat{\Lambda}_0 = \hat{\Lambda}_i = 0 \tag{55}$$

and the hydrodynamic variables transform as

$$\rho' = \rho\left[1 - \varepsilon 3\left(\tfrac{1}{2}\beta + \alpha t\right)\right], \quad v'_i = v_i + \varepsilon\left[\omega_{ij}v_j + \alpha x_i - u_i - \left(\tfrac{1}{2}\beta + \alpha t\right)v_i\right],$$

$$S' = S + \varepsilon m\left(\tfrac{1}{2}\alpha x_i x_i - u_i x_i\right). \tag{56}$$

For a given Lagrangian-picture symmetry, the sets of functions $(\bar{P},\bar{J}_i)$ and $(\hat{P},\hat{J}_i)$ do not always coincide; an example is given in the next section. Note that the Lagrangian (51) is not directly connected with the Lagrangian (4); their relation is examined in [6].

**5. Pure relabelling symmetry**

In (45), choose $\xi_0 = \eta_i = 0$. The non-trivial component of the transformation reduces to $a'_i = a_i + \varepsilon\xi_i(a)$, which corresponds to the infinitesimal form of the pure label transformation described in Sec. 3. To see the significance of the constraint (46) we consider the infinitesimal transformation of the reference density, which is, in general,

$$\rho'_0(a') = \rho_0(a) + \varepsilon\delta\rho_0(a) + \varepsilon\frac{\partial\rho_0}{\partial a_i}\xi_i(a,t) \tag{57}$$

where $\delta\rho_0(a) = \rho'_0(a) - \rho_0(a)$ is the functional variation. The infinitesimal form of (35) is then

$$\delta\rho_0 + \frac{\partial}{\partial a_i}(\rho_0\xi_i) = 0. \tag{58}$$

Hence, comparing with (46), we see that the latter constrains $\xi_i$ so that $\rho_0$ is an invariant function ($\delta\rho_0 = 0$). Note that this constraint generally prevents $\xi_i$ from being chosen constant.

The conserved density and current associated with the relabel symmetry are

$$\mathrm{P}(a,t)=-m\rho_0\frac{\partial q_i}{\partial t}\frac{\partial q_i}{\partial a_j}\xi_j, \quad J_i(a,t)=\rho_0\xi_i\left(\tfrac{1}{2}m\frac{\partial q_i}{\partial t}\frac{\partial q_i}{\partial t}-V(q(a,t))-V_Q\right). \tag{59}$$

This conservation law is known in the analogous classical fluid theory [20]. An interesting feature of it is that the term in brackets in $J_i$ is the Lagrangian of a particle of mass $m$ moving in the potential $V+V_Q$. This suggests that this Lagrangian may acquire a physical significance as a component of the current.

To translate these expressions into the corresponding Eulerian quantities we define the function $\bar{\xi}_i(x,t)=\left(\partial q_i/\partial a_j\right)\xi_j\big|_{a(x,t)}$ so that the density and current (59) become

$$\bar{\mathrm{P}}(x,t)=-m\rho v_i\bar{\xi}_i, \quad \bar{J}_i(x,t)=\rho\bar{\xi}_j\left[-mv_iv_j+\delta_{ij}\left(\tfrac{1}{2}mv^2-V-V_Q\right)\right]. \tag{60}$$

To obtain the conditions obeyed by $\bar{\xi}_i$, we note that the label is a constant of the motion along the trajectory it defines, so that, regarded as a function of the Eulerian independent variables, i.e., $a_i(x,t)$, it satisfies

$$\frac{\partial a_i}{\partial t}+v_j\frac{\partial a_i}{\partial x_j}=0. \tag{61}$$

Any function $\xi_i$ of $a_i$ is also a constant of the motion and obeys the same equation. Translating the latter and (46) into the Eulerian language, the function $\bar{\xi}_i$ obeys the two relations

$$\frac{\partial\bar{\xi}_i}{\partial t}+v_j\frac{\partial\bar{\xi}_i}{\partial x_j}=\bar{\xi}_j\frac{\partial v_i}{\partial x_j}, \quad \frac{\partial}{\partial x_i}\left(\rho\bar{\xi}_i\right)=0. \tag{62}$$

The transformation generates an infinite class of conservation laws depending on the choice of $\xi_i$ (subject to (46)), and encapsulates known laws as special cases. One noteworthy aspect is that this vector projects out of the linear momentum density a function that is conserved even when $\partial V/\partial q_i\neq 0$. As an example of the application of this symmetry, we shall derive from it Kelvin's theorem of the conservation of circulation for the quantum fluid, following a method employed in an analogous classical example [14]. Denote by $C_0$ a closed loop $a_i(s)$ in label space parameterized by the arc length $s$ and define the transformation function to be

$$\xi_i(a)=(1/\rho_0)\oint_{C_0}\delta(a_i-a_i(s))da_i(s). \tag{63}$$

This function is easily seen to obey (46). Substituting in (44) we obtain

$$\frac{d}{dt}\oint_{C_0}\frac{\partial q_i(a(s),t)}{\partial t}\frac{\partial q_i(a(s),t)}{\partial a_j}da_j(s)=0. \tag{64}$$

Changing variables from label to current position, the integration will be over a closed loop $C$ defined by the instantaneous positions at time $t$ occupied by the particles that originally comprised the circuit $C_0$ [25,26]. Then

$$\frac{d}{dt}\oint_C \frac{\partial q_i(a(s),t)}{\partial t} dq_i(s,t) = 0 \tag{65}$$

which is just (20). We conclude that Kelvin's theorem in quantum theory is a consequence of invariance with respect to the relabelling group.

The corresponding Eulerian transformation (52) is the identity with no useful associated conserved quantity $\hat{P}$.

## 6. Spacetime symmetries. The Schrödinger group

Here we shall dissect the 12-parameter subgroup of (45), obtained by setting $\xi_i = 0$, characterizing each case by a subset of non-zero parameters.

(i) *Time translation*: $d \neq 0$. Then $\partial V/\partial t = 0$ and the conserved Lagrangian and Eulerian densities are

$$P = -dH(a,t), \quad \bar{P}(x,t) = \hat{P}(x,t) = -d\bar{H}(x,t) \tag{66}$$

where

$$H = \tfrac{1}{2} m\rho_0 \frac{\partial q_i}{\partial t}\frac{\partial q_i}{\partial t} + \rho_0 U + \rho_0 V, \quad \bar{H} = HJ^{-1}\big|_{a(x,t)} = \tfrac{1}{2} m\rho v^2 + \rho U + \rho V \tag{67}$$

is the energy density.

(ii) *Space translation*: $c_i \neq 0$. Then $\partial V/\partial q_i = 0$ and

$$P = c_i m\rho_0 \frac{\partial q_i}{\partial t}, \quad \bar{P} = \hat{P} = c_i m\rho v_i \tag{68}$$

is proportional to the linear momentum density.

(iii) *Space rotation*: $\omega_{ij} \neq 0$. Then $\varepsilon_{ijk} q_j(\partial V/\partial q_k) = 0$ and

$$P = \omega_{ij} m\rho_0 q_j \frac{\partial q_i}{\partial t}, \quad \bar{P} = \hat{P} = \omega_{ij} m\rho q_j v_i \tag{69}$$

is proportional to the angular momentum density.

(iv) *Galilean boost*: $u_i \neq 0$. Then $\partial V/\partial q_i = 0$ and

$$P = u_i m \rho_0 \left( q_i - t \frac{\partial q_i}{\partial t} \right), \quad \bar{P} = \hat{P} = u_i m \rho (x_i - t v_i) \tag{70}$$

is proportional to the Galilean momentum density.

(v) *Dilation*: $\beta \neq 0$. Then

$$q_i \frac{\partial V}{\partial q_i} + 2t \frac{\partial V}{\partial t} + 2V = 0 \tag{71}$$

and

$$P = \beta \left( \tfrac{1}{2} m \rho_0 q_i \frac{\partial q_i}{\partial t} - tH \right), \quad \bar{P} = \hat{P} = \beta \left( \tfrac{1}{2} m \rho x_i v_i - t\bar{H} \right). \tag{72}$$

(vi) *Extension*: $\alpha \neq 0$. Then

$$q_i \frac{\partial V}{\partial q_i} + t \frac{\partial V}{\partial t} + 2V = 0 \tag{73}$$

and

$$P = \alpha \left[ m \rho_0 q_i \left( t \frac{\partial q_i}{\partial t} - \tfrac{1}{2} q_i \right) - t^2 H \right], \quad \bar{P} = \hat{P} = \alpha \left[ m \rho x_i (t v_i - \tfrac{1}{2} x_i) - t^2 \bar{H} \right]. \tag{74}$$

In all cases the two Eulerian densities defined in (49) and (53) are the same. Cases (i), (ii) and (iii) correspond to the usual energy, momentum and angular momentum continuity equations for the Schrödinger field [17]. In case (iv) the continuity equation reduces to ($t$ times) the Euler-Lagrange equations (11) (subject to vanishing external force), which, as can be checked, indeed have the form (41). These four cases together correspond to the 10-parameter Galilean group. They are supplemented by two further one-parameter kinematical transformations corresponding to cases (v) and (vi). These transformations were originally discovered as symmetries of the free Schrödinger equation [27] and the complete 12-parameter set defines the 'Schrödinger group'. Subsequently, it was realized that dilation and extension are also symmetries for a non-trivial class of potentials [28,29]. The two classes of admissible potentials we have given in (71) and (73) generalize those stated in [29], where it is assumed that the transformations are applied simultaneously. These additional symmetries were not found in the analogous prior work on a classical fluid [20] although it is known that the Schrödinger group is the maximal covariance group of classical hydrodynamics [30].

### 7. Linear superposition as a relabelling transformation

In this work we have been concerned with transformations under which the dynamical equations of quantum hydrodynamics are covariant. These symmetries provide a method of building new solutions from known ones. This perspective suggests treating the linear superposition of wavefunctions, which achieves the same constructive end, as a type of symmetry, i.e., a transformation with respect to which the Schrödinger equation is covariant. In this case the old and new solutions are, in general, not physically equivalent.

Consider two arbitrary wavefunctions $\psi$ and $\phi$ which are both finite at the spacetime point $(x,t)$. Their linear superposition results in a new solution also finite at the point: $\psi' = \psi + \varepsilon\phi$. We shall regard this relation as a transformation of $\psi$ in which the transformation function $\phi$ also obeys the Schrödinger equation, restricting attention to infinitesimal transformations where the real parameter $\varepsilon$ is chosen so that $|\varepsilon\phi| \ll |\psi|$ (we can generalize so that $\varepsilon$ is complex but will not do so). Under this transformation the independent and dependent variables of the Eulerian picture transform as

$$t' = t, \quad x'_i = x_i, \quad \psi'(x',t') = \psi(x,t) + \varepsilon\phi(x,t). \tag{75}$$

To see the value of this alternative view of superposition, we shall derive the conserved charge corresponding to this symmetry implied by Noether's theorem. Using the fact that $\psi$ and $\phi$ both obey the wave equation, the Lagrangian density (51) transforms under (75) as

$$\hat{\ell}(\psi',\partial\psi'/\partial t',\partial\psi'/\partial x') = \hat{\ell}(\psi,\partial\psi/\partial t,\partial\psi/\partial x) - \varepsilon\frac{\hbar^2}{4m}\frac{\partial^2}{\partial x_i \partial x_i}(\psi^*\phi + \phi^*\psi). \tag{76}$$

The transformation is therefore Noetherian, i.e., it leaves the action invariant. The Noether charge is $\int (i\hbar/2)(\psi^*\phi - \phi^*\psi)d^3x$. Thus, with respect to the Lagrangian (51), the superposition symmetry is correlated with (the real part of) the scalar product of the solutions $\psi,\phi$, which is indeed conserved by the Schrödinger equation.

Choosing $\phi = -i\psi$, this approach contains as a special case an infinitesimal gauge transformation where the corresponding charge is $\int \hbar\psi^*\psi d^3x$, as expected.

The corresponding transformation of the independent and dependent variables in the Lagrangian picture is:

$$t' = t, \quad a'_i = a'_i(a,t), \quad q'_i(a',t') = q_i(a,t). \tag{77}$$

In the Lagrangian picture we may distinguish the wavefunctions under consideration at given $x$ and $t$ by the labels of the corresponding unique trajectories that generate them. We may then characterize the process of superposition as a transformation of label. Letting $a_i$ label $\psi$, we seek the label $a'_i$ of the unique trajectory passing the point $x$ at time $t$ that generates $\psi'$. Given the initial wavefunction $\psi'_0$, we can solve (11) to find the trajectory $q'_i(a',t',\psi'_0)$. Inverting, we have $a'_i(q',t,\psi'_0)$ and substituting $q'_i(a',t',\psi'_0) = q_i(a,t,\psi_0)$ gives $a'_i = a'_i(a,t,\psi_0,\phi_0)$.

In the method just described the new label is computed by first solving for the transformed trajectories. We now consider a special case where the transformed label may be computed from the original trajectories by supposing that the infinitesimal term in (75) is obtained by varying a typical continuously variable parameter $A$ (assumed dimensionless) on which the function $\psi$ depends. For example, if $\psi$ is a packet function, $A$ could be proportional to its initial width. Let $\psi'$ correspond to $\psi$ evaluated for $A' = A + \varepsilon$. Then

$$\psi'(x,t,A') = \psi(x,t,A) + \varepsilon \frac{\partial \psi}{\partial A}. \tag{78}$$

The Hamiltonian is independent of $A$ so the function $\partial \psi / \partial A$ is a solution if $\psi$ is. The Eulerian fields transform as

$$\rho'(A') = \rho(A) + \varepsilon \frac{\partial \rho}{\partial A}, \quad v_i'(A') = v_i(A) + \varepsilon \frac{\partial v_i}{\partial A}. \tag{79}$$

To determine the label $a_i'$ that generates $\psi'$ we observe that the time-dependence of $\partial \psi / \partial A$ can be computed from the trajectories in two ways: by propagating $\partial \psi_0 / \partial A$, or by derivating having found $\psi$ by propagating $\psi_0$. From the latter we expect that $a_i'$ can be expressed just in terms of $a_i$ and $\psi_0$. We have

$$t' = t, \quad a_i'(a,t,A) = a_i + \varepsilon \xi_i(a,t,A), \quad q_i'(a',t',A') = q_i(a,t,A). \tag{80}$$

A straightforward way to calculate $\xi_i$ is to note first that $q_i$ is an invariant function of $A$ and $a_i$. Then

$$q_i'(a',t,A') = q_i(a',t,A')$$
$$= q_i(a,t,A) + \varepsilon \frac{\partial q_i}{\partial a_j} \xi_j + \varepsilon \frac{\partial q_i}{\partial A}. \tag{81}$$

Combining this with the last member of (80) gives

$$\frac{\partial q_i}{\partial a_j} \xi_j + \frac{\partial q_i}{\partial A} = 0. \tag{82}$$

Inverting the deformation matrix using (8) thus gives finally

$$\xi_j = -J^{-1} J_{ij} \frac{\partial q_i}{\partial A}. \tag{83}$$

To check that this transformation is a symmetry of the trajectory law (11), we only need confirm consistency between the transformation rules of the Lagrangian velocity and density implied by the first two members of (39) (with (83) substituted) and the Eulerian rules (79). This follows straightforwardly using the following formula for conversion to the Lagrangian picture, which holds for any Eulerian function $f(x,t,A)$:

$$\frac{\partial}{\partial A}\left[ f(x,t,A)\big|_{x=q(a,t,A)} \right] = \frac{\partial f(x,t,A)}{\partial A}\bigg|_{x=q(a,t,A)} + \frac{\partial f(x,t,A)}{\partial x_i} \frac{\partial x_i}{\partial A}\bigg|_{x=q(a,t,A)}. \tag{84}$$

Inverting the procedure that led to the function (83), we may assert that *the linear superposition of wavefunctions (78) may be generated by the following infinitesimal deformation-dependent relabelling of the trajectories, a symmetry of the law of motion (11)*:

$$t' = t, \quad a'_i(a,t,A) = a_i - \varepsilon J^{-1} J_{ji} \frac{\partial q_j}{\partial A}, \quad q'_i(a',t',A') = q_i(a,t,A). \tag{85}$$

This transformation is consistent with the label transformation found in (45) since it involves the deformation coefficients, a dependence that was excluded in the analysis of Sec. 4. It is evident that we may generalize the representation of linear superposition by a particle-relabelling transformation to the most general case, where $\phi$ is any solution of the Schrödinger equation.

**8. Conclusion**

Although it has proved fertile in computational quantum chemistry, the development of the Lagrangian picture of quantum mechanics has been uneven; key formal aspects of the picture that could have an impact in the numerical endeavour have yet to be fully investigated. Here we have explored aspects of the Lagrangian picture that provide an alternative perspective on symmetries and conservation laws in quantum mechanics. Pursuing a variational technique, we have derived the maximal kinematical covariance group of the Schrödinger equation and the associated conserved charges, and established connections between the Lagrangian method and the known Eulerian description. A novel aspect is the appearance of an infinite parameter relabelling group, which implies a wide set of conservation laws such as Kelvin's circulation theorem. Moving beyond the variational method, we have developed an alternative perspective on the linear superposition principle as a relabelling symmetry.

In our variational approach we restricted the possible functional dependence of the transformation functions. This is sufficient to generate all the kinematical symmetries of the Schrödinger equation, together with relabellings, but future work might allow dependence of the transformation functions on the deformation coefficients. In addition, an examination of alternative Lagrangians could prove valuable. On the other hand, discovering conservation laws via the variational approach has limitations and direct methods may be more efficient [21].